# Exploring phase transitions and thermal dynamics in nanoconfined liquid crystals using liquid-phase TEM

Olga Kaczmarczyk[1], Konrad Cyprych[2], Dominika Benkowska-Biernacka[1], Rafał Kowalczyk[3], Katarzyna Matczyszyn[1], Hanglong Wu[4], Frances M. Ross[4], Andrzej Miniewicz[1], Andrzej Żak*[1,4]

[1] Institute of Advanced Materials, Wroclaw University of Science and Technology, 50-370 Wroclaw, Poland

[2] Soft Matter Optics Group, Faculty of Chemistry, Wroclaw University of Science and Technology, 50-370 Wroclaw, Poland

[3] Department of Bioorganic Chemistry, Faculty of Chemistry, Wroclaw University of Science and Technology, 50-370 Wroclaw, Poland

[4] Department of Material Science and Engineering, Massachusetts Institute of Technology, Cambridge, Massachusetts 02139, United States

* andrzej.zak@pwr.edu.pl

**Abstract**

Nanoconfined liquid crystals (LCs) and their nanocomposites are driving the next generation of photonic applications. Consequently, deepening our understanding of mesophase stability, defect topology, and the dynamic response of LCs at the nanoscale requires the development of novel characterization approaches. This motivates us to perform in situ observations on model 4'-octyl-4-cyanobiphenyl (8CB) LC using liquid-phase scanning transmission electron microscopy (LP-STEM). We find that the electron beam induced consecutive phase changes from smectic A to nematic (SmA-N) and from nematic to isotropic (N-I). The kinetic dependence of the phase transition on dose rate shows that the time between SmA-N and N-I shortens with increasing rate, revealing the hypothesis that a higher electron dose rate increases the energy dissipation rate, leading to substantial heat generation in the sample. We report on the spontaneous formation of disclinations, ordering effects, and complete process reversibility. Radiolytic effects of the electron beam are discussed in detail, and additional experiments with external heating indicate that the observed phenomena are mainly thermal in nature. The results are supported by calculations of heat diffusion, suggesting the nanoconfined 8CB differs significantly in thermal properties compared to the bulk one. This is the first detailed study of LC phase transitions using LP-STEM, which paves the way for further studies of nanoconfined LCs and for the development of the technique for advanced LC materials research.

**Introduction**

Liquid crystals (LCs)[1] are known for their unique anisotropic properties and the ability to flow while maintaining well-defined ordered structures. Each mesophase exists within specific temperature range or within certain combination of temperature and concentration, depending on whether the LC is thermotropic or lyotropic, respectively. LCs can respond to external signals, e.g. temperature[2], electric and magnetic fields[3,4], light[5,6], surface interactions[7], and mechanical stresses[8], which enables precise manipulation of their optical and dielectric properties [9]. These features prove them to be highly suitable for different applications, including displays[10], photonic devices[11–13], modern advanced and smart materials technologies[14–16], sensing[17], and even potential usage in quantum computing[18] and nanomedicine[19–21].

LC-forming molecules are typically polar, possessing a permanent electric dipole moment[1] and rod- or disc-like shapes. Among the thermotropic LCs, a widely studied rod-shaped mesogene is 4'-octyl-4-cyanobiphenyl (8CB)[22–24]. The 8CB molecule has a permanent dipole moment of $\mu = 6.2$ D and consists of a cyano (CN) group attached to biphenyl core and a long alkyl chain[24]. 8CB molecules, 1.8 nm in length, tend to form dimers by the π-π stacking of adjacent biphenyls, resulting in a smectic bi-layer of 3.2 nm in thickness. This layer is shorter than the sum of the length of the two individual 8CB molecules, because the layers are interdigitated, with molecules from one layer partially penetrating into the next [25–27]. 8CB is found to form two primary mesophases, e.i. smectic A (SmA or more precisely $SmA_d$ – the dimeric smectic A) and nematic (N), with relatively low phase transition temperatures below 40°C[23]. Excluding the crystallization process, which exhibits significant thermal hysteresis (a temperature shift between heating and cooling), the transition temperatures for SmA-N and N-I show negligible changes during slow heating and cooling cycles[23]. In nanoconfined 8CB LC, the phase-transition temperatures could differ from those observed in the bulk sample [23], being lower for the nanometer-thick layer [28–31].

The conventional methods used for LC characterization are spectroscopy[32], x-ray diffraction (XRD)[33], polarized light microscopy (PLM)[34], differential scanning calorimetry (DSC)[23], small-angle x-ray scattering (SAXS)[35,36], and nonlinear optical (NLO) techniques[37,38]. (Scanning) transmission electron microscopy ((S)TEM) has also been used for nanoscale LCs characterization. Cryo-TEM and thin-sections imaging of LC are well-established methods that allow for observing specific mesophases with high spatial resolution[39–41]. However, this cannot be used for studying the phase transition sequence in LCs because of the static nature of thin sections and vitrified samples. A few reports highlight the application of other electron microscopy techniques. For instance, 4D-STEM[42] has been utilized for columnar LC phase growth from vapor-deposited glassy material[43,44], and scanning electron microscopy (SEM) was used for director field visualization in nematic LC polymers[45]. Nevertheless, in-situ imaging of liquid-crystalline materials is necessary for a profound understanding of the processes associated with topological defects formation and changes in phase behavior with temperature [46], their application as templates for molecular self-assembly[47], and, especially, LC interactions with nanomaterials at the nanoscale for the development of novel metamaterials [48].

Thin layers and nanoconfined LCs are particularly interesting because of their potential applications in modern opto-electronics[49] as photonic crystals[50] and thin-film transistors[51]. A promising technique for studying thin layers is liquid phase-(S)TEM (LP-(S)TEM), which yields information on liquid phase dynamics in fields of science including nanotechnology, biology, nanomedicine, and electrochemistry[52–65]. LP-(S)TEM *in situ* studies involving the liquid crystalline phase have been reported only once in the literature[66] based solely on electron diffraction analysis. The ability to record images, and hence spatially and temporally resolved information on LC phase transformations, is an exciting prospect provided that the radiation effects on the material are considered in interpretating the results[52,63–65,67–70].

In this work, we studied phase changes, nucleation sites, defect formation and surface tension effects near N-I transitions in nanoconfined 8CB using LP-STEM. Our measurements of phase transformation kinetics strongly suggest that temperature rise due to irradiation is the driving force of the phase transitions, and temperature increases with the observation time and cumulative electron dose. Thermal changes in LP-(S)TEM imaging in liquid samples are still poorly understood. The exact temperature rise is still challenging to measure in such small volumes in the order of several nanoliters[71], and in most cases of aqueous solutions, the temperature increase could be neglected[62,64]. Our findings show that the temperature increase in the LC sample can be much higher than that expected in aqueous solutions, enabling a better understanding of the thermal processes in STEM of nonaqueous liquid samples . This paper is divided into three sections. First, we present LP-STEM imaging of LC and describe the processes of phase transitions, defect formation, and surface effects. Then, we discuss the thermal origin of the observed phase changes supported by heat diffusion calculations. The thermal and radiolysis effects are discussed based on control experiments with additional laser heating, and via process reversibility. The paper concludes with a discussion of the contrast seen in the images and the perspectives this implies for LP-STEM imaging of LC materials. The results advance our fundamental understanding of LCs and offer insight into how LP-STEM can be used in further research on LC materials, emphasizing their relevance to nanocomposites and novel functional materials. Finally, manipulating the structure of a liquid crystal by an electron beam suggests the prospect of control at the nanoscale and design of advanced materials.

**Results and discussion**

*Defect formation, surface tension effects and phase transition kinetics of 8CB under various electron dose conditions*

Figure 1 illustrates the temperature-dependent structures of 8CB, along with the results of imaging the material using electrons at various dose rates. The LC sample was prepared by sandwiching 8CB between two standard TEM Cu grids coated with approximately 25 nm thick amorphous carbon film[72–74]. This sample preparation procedure is presented in Fig.1 b with a polarizing microscope image of the sandwiched sample in Fig. S1, and more preparation details in Methods. This preparation method does not ensure uniform LC thickness or full isolation from the vacuum throughout the sample, and this should be considered when estimating temperature changes associated with specific phase transitions. In the high vacuum environment the phase change temperatures of the LC sample may be shifted, and these conditions are in any case far from the standard LC application. Moreover, the LC needs to be enclosed on both sides to achieve a thin layer. Having the same support on top and bottom provides uniform effects on the thin film sample, which is crucial for correlating two-dimensional STEM imaging with the actual three-dimensional structure. The thickness of a typical observation area was 20-40 nm, estimated using the method described in Supplementary Information (Fig. S3 and S4). For such a low LC thickness, the N-I phase transition may shift to lower temperatures and the surface effects such as anchoring may force LC alignment in a specific direction[28,31,75]. As the LC molecules may also reorient in electric and magnetic fields[76,77], the imaging was performed in low-field STEM conditions in which the magnetic field from the objective lens, which is typically in the range of 1.5-2 T, was lowered to 0.12 T to minimize the interaction with the LC. . The magnetic field value for Fréedericksz transition of 8CB depends on sample thickness and temperature, being relatively high (0.1- 1 T) due to low magnetic susceptibility of LCs molecules[78,79]. For such an organic sample with a relatively low molecular mass, the electron scattering is relatively small. Consequently, we used low magnification STEM (LM-STEM) at the shortest available camera length , as described in the Methods section. Despite the choice of this detector, our images exhibited contrast characteristics typical of an annular bright field (ABF) image, meaning that the thicker and more scattering features appear darker in the image. The observations were performed at several probe currents to check the mesophase behavior in LP-STEM at different electron dose rates, with four different dose rates of 2.72, 1.49, 0.79, and 0.25 e$^-$/nm$^2$s. For each dose rate, a separate observation area was scanned (Fig. S2a-c). All areas imaged were close to each other to minimize the thickness differences. The imaging details are provided in Methods

We found that the electron beam induced a sequence of very similar processes for the first three electron dose rates. By tracking the contrast changes and process behavior, we were able to recognize a few crucial phenomena: SmA-N and N-I phase transitions, the resulting N and I phases growth and propagation, and defect formation. The defects in LCs, also called disclinations, are a local discontinuities in the material's ordered structure[80]. Defects can be treated as areas of lower density isotropic phase and can form specific structures utilized in technology[46,47,80,81].

At 25 °C, the material is in the SmA phase, which does not give a strong contrast in STEM, appearing mostly homogeneous through the observation area (first row in Fig. 1cd). The onset and subsequent growth of a new phase are evidenced by the darker spots that appear randomly throughout the observation areas after 23 s of irradiation with 2.72 and 1.49 e$^-$/nm$^2$s, and after 26 s at 0.79 e$^-$/nm$^2$s (second row in Fig. 1cde). With further electron irradiation, the phase spreads uniformly across the observation area (Supplementary videos 1, 2, and 3, available at data repository[82]). Such uniform nucleation behavior and homogenous "spreading" of the new phase is characteristic of the second-order or very weak first-order phase transition, typically observed in LCs SmA-N phase transitions[23,83]. Therefore, the latest, darker phase is assumed to be N mesophase, which in bulk exists typically around 32.7°C[23]. Upon further electron irradiation, the N phase domains grow rapidly, leading to elongated disclinations at domain boundaries (third row in Fig. 1cde). This defect follows the N phase propagation direction and narrows to about 250 nm in width. Further electron irradiation led to the formation of another, less scattering phase, which nucleates at the defect (fifth row in Fig. 1cde) at different times. The new phase shows the same contrast as the disclination, suggesting that their properties are similar. Since a disclination can be treated as an area of a disordered, isotropic liquid, our observed phase transition is consistent with expectations that the defects serve as a pre-existing nucleation center of the disordered phase, at which the energetic cost of isotropic phase growth is lower than in pure N mesophase. Therefore, the second phase transition observed in this experiment is identified as N-I, occurring around 39.9°C in a bulk 8CB[23]. Although the time to this N-I transformation differs with electron dose rate, the cumulative doses were similar, at 87, 77, and 83 e$^-$/nm$^2$, respectively. After the N-I transformation started at the defect position, the further nucleation of the I phase appeared at localized regions and grew quickly (the last row in Fig. 1cde). The nucleation behavior of this phase roughly suggests the first-order type of phase transition, which is typical for N-I[84].

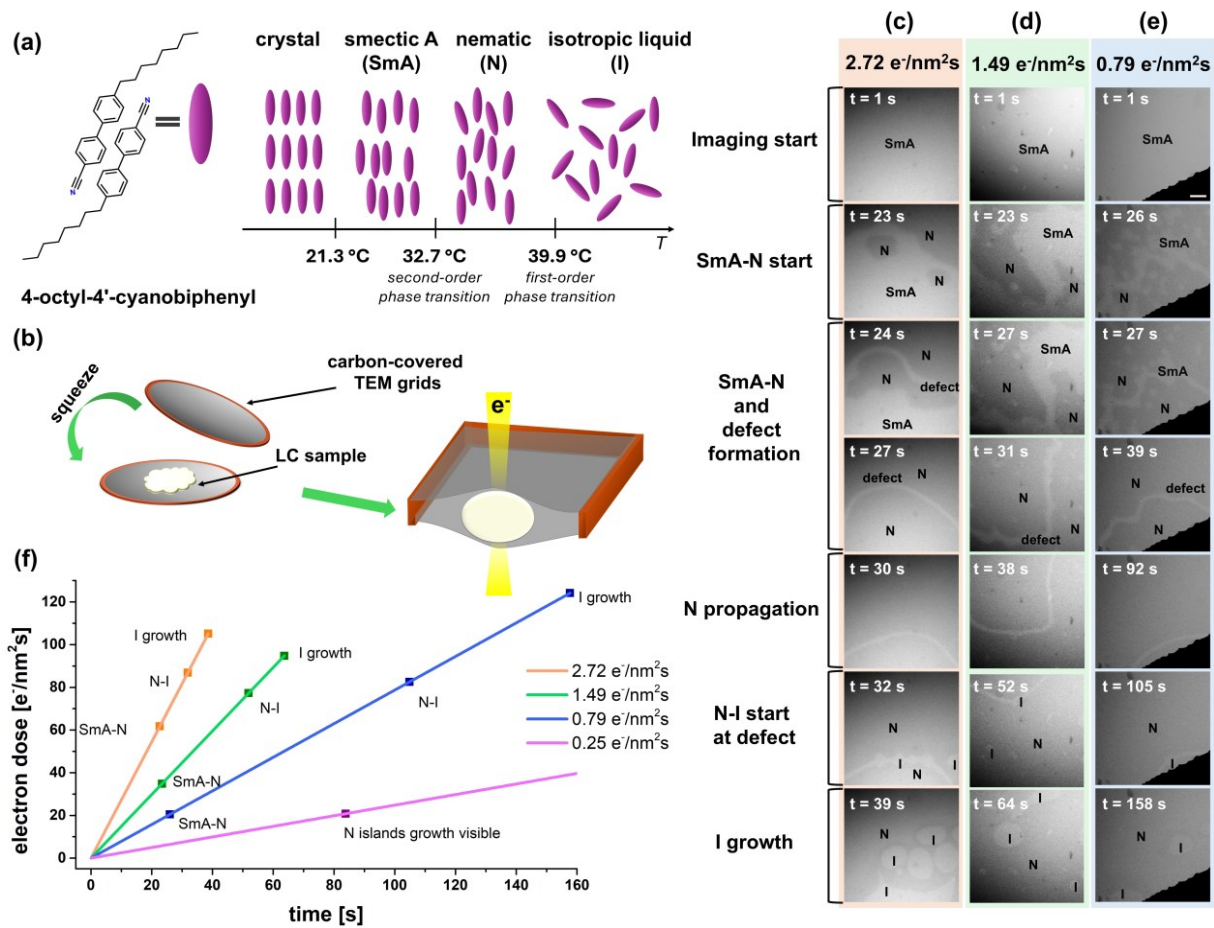

**Figure 1**. Graphical representation of 8CB LC molecules and mesophases with phase changes temperatures (a) and sample preparation procedure with liquid-cell schematic cross section (b). In SmA phase 8CB dimers form the layers, which are interdigitated, with molecules from one layer partially penetrating into the next. This is specific only to SmA phase, whereas in N mesophase the molecules have more freedom and do not form dimers and layers. LP-STEM of 8CB SmA-N and N-I phase changes at four electron dose rates : (c) 2.72; (d) 1.49; (e) 0.79 e$^-$/nm$^2$s. The results for dose rate of 0.25 e$^-$/nm$^2$s are provided in the Fig. S2d in Supplementary Information. For the each column (c-e), the same processes occur in each row: SmA mesophase, SmA-N transition, SmA-N continuation and defect formation, N mesophase propagation, N-I transition start at defect, and further growth of I phase. The phase transition times and corresponding cumulative electron doses are plotted in panel (f). The end points denoted as "I growth" indicate the electron doses at which the images in the last columns of panels (c-e) were acquired. For the full-time series of processes occurring in (c-d) see supplementary videos 1-3[82]. The scale bar (first row in e) is the same for all images and represents 2 μm.

In the case of the lowest studied electron dose rate of 0.25 e$^-$/nm$^2$s, N islands start to grow after about 84 s of electron irradiation at a cumulative dose of 21 e$^-$/nm$^2$, although the exact determination of the time is difficult due to the shorter dwell time and greater noise in the images (Fig. S2d). During the whole time of observation, 1586 s (26 min), no full set of phase transitions took place and only N islands expansion was observed. The final cumulative electron dose reached 393 e$^-$/nm$^2$, which was much higher than in the higher electron dose rate experiments, but of course spread over greater time.

The electron dose versus time analysis of phase transition temperatures (Fig. 1f) shows a trend: with increasing electron dose rate, the time between the SmA-N and N-I transitions becomes shorter. The SmA-N transition starts at a similar time across all rates, and N-I occurs at a similar cumulative dose. We propose that this effect is connected to sample heating due to interaction with electron beam, and discuss this in the following paragraphs.

In the case of the lowest studied electron dose rate of 0.25 e-/nm2s, N islands start to grow after about 84 s of electron irradiation at a cumulative dose of 21 e-/nm2, although the exact determination of the time is difficult due to the shorter dwell time and greater noise in the images (Fig. S2d). During the whole time of observation, 1586 s (26 min), no full set of phase transitions took place and only N islands expansion was observed. The final cumulative electron dose reached 393 e-/nm2, which was much higher than in the higher electron dose rate experiments, but of course spread over greater time.

We show in Fig. 2 an example of the analysis of the phase transformation in a uniform region under the influence of the electron beam, occurring when the material transforms from the N to the I phase. The homogenous area was irradiated with a constant electron dose rate of 2.72 e$^-$/nm$^2$s until the N phase covered most of the studied region (Fig. 2a). The corresponding intensity profile of pure N phase is flat, with slight variations probably linked to minor changes in sample thickness (Fig. 2a, below the STEM image). With increasing exposure to the electron beam, the nucleation of I phase begins in the center of the observation area (Fig. 2b). Minor visual changes in brightness in the STEM image are visible in the corresponding intensity profile plot. This change represents the first nucleation of I islands, with a size of several dozen nanometers. The small peak in the intensity profile plot for Fig. 2b can be also related to another phenomenon, better visible in the following image in Fig. 2c. Once the nucleation of I phase begins at the nanoscale, its further growth in an oval shape happens quickly (Fig. 2c). Around the I phase island a lighter ring appears and corresponding increase in the intensity profile plot for distances ranges 1.5 – 4 μm and 22 – 25 μm is evident (*cf*. orange-dashed line squares in Fig. 2c). With further propagation of the I phase (Fig. 2d), these effects become less visible. This effect may be connected to the known phenomenon of pre-transitional surface ordering in LCs[85,86]. Unlike conventional liquids, LC materials exhibit a unique property of increasing surface ordering near the phase transition temperature (especially N-I), thereby locally increasing surface tension (the temperature coefficient of surface tension becomes positive with increasing temperature, an unusual feature). Therefore, these local order-parameter fluctuations may lead to changes in STEM contrast – the areas closest to the I phase show the highest contrast (a sudden decrease in the intensity profile), indicating higher ordering in these areas. Additionally, this increase in surface tension at the N- I phase boundary forces the I phase to grow in a spherical shape.

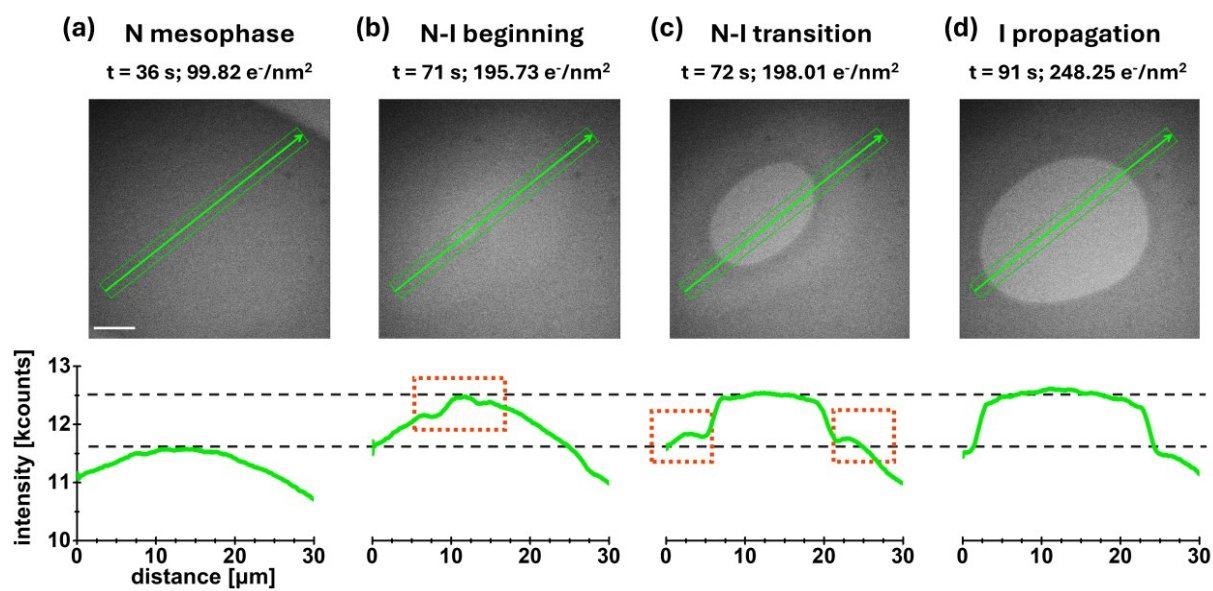

**Figure 2**. LP-STEM of N-I phase transformation with corresponding intensity profile plots below. The arrow indicates the contrast profile plot direction, and the outline shows the integration width. (a) Homogenous N mesophase propagates with flat intensity profile plot. (b) The beginning of the N-I transition. A lighter area in the center can be observed, corresponding to the significant increase in the contrast line profile. This change is probably related to nucleation of the I phase in the nanoscale and order parameter fluctuations in the pre-transitional temperature (1 second before significant changes occurred). (c) Direct N-I transition. An oval island of I phase appears in the center of the imaging area, which results in the broader peak in the intensity distribution. The surrounding brighter ring is probably the effect of order parameter fluctuations and is represented as smaller peaks in the contrast plot (marked with orange frames). (d) Further growth of I phase. The corresponding plot mainly represents bright I phase without visible changes in contrast near the phase boundary. The electron dose rate used in this experiment was 2.72 e-/nm$^2$s. A whole series of images of the process is presented in Supplementary video 4[82]. The scale bar is 5 μm.

We now discuss the results shown in Fig. 1 and Fig. 2 based on the hypothesis that a higher electron dose rate increases the energy dissipation rate, resulting in substantial heat generation on the sample over a short period. The SmA-N and N-I phase changes occur in the bulk material at around 32.7°C and 39.9°C, respectively[23]. With an initial temperature within the TEM polepiece assumed to be 25.0 °C, our experiments showed the sequence of phase transitions, that is expected to occur at temperature differences of approximately 7.9 °C (SmA-N) and 7.0 °C (N-I)[23]. This suggests an overall temperature increase of at least ΔT > 15.0 °C within less than 40 seconds of scanned electron irradiation at the highest electron dose rate used in this study (2.72 e-/nm²s) (*cf*. Fig. 1c, last column). In the case of nanoconfined 8CB, the N-I phase transition temperature can be lower, but, according to the literature, for material confined in 25 nm pores the shift is in a range of only 1°C[28].

Such a high temperature increase raises concerns regarding literature reports, that indicate that typical LP-STEM imaging yields a negligible rise of a few degrees maximum[62–64]. However, most of these estimations were done for aqueous solutions. In our case, we are investigating molecules quite different from H$_2$O. LCs exhibit notable differences in physical and chemical properties compared to water, which could certainly affect sample heating (see the comparison in Table S1). Regarding its molecular properties, 8CB has a relatively complex structure consisting of a π-electron rich biphenyl core, a polar cyano group (-CN), and a long flexible alkyl chain (C$_8$H$_{17}$). Typically, the LC molecule has more electrons, and a biphenyl core adds delocalized π electrons to the structure. These electrons increase the inelastic scattering of the electron beam, as reported in literature by measuring electron energy-loss spectroscopy (EELS) of aromatic compounds[87]. The generally higher inelastic scattering of electrons in organic compounds as compared to water can also be deduced from the electron inelastic mean free paths (IMFPs) calculations presented in the literature[88]. For 200 keV (the value consistent with our experiments) the calculated IMPFs are smaller for organic compounds with values in the range of 159-209 nm, compared to 238 nm for water[88]. As a result, in organic compounds, there may be a greater temperature increase than in water due to higher inelastic scattering. Unfortunately, no specific value for 8CB LC is available in the literature, but the overall trend can be inferred. 8CB and water also notably differ in thermal properties like thermal conductivity, specific heat capacity, and thermal diffusivity (Table S1). In general, 8CB distributes the heat more slowly than water, leading to a higher temperature gradient within the irradiated area because of lower thermal conductivity. This suggests that LC material may be affected by the electron beam differently compared to water, indicating that additional factors should be considered when estimating the temperature rise during LP-STEM.

Under any assumptions, we expect the lower electron dose rate to result in a smaller temperature gradient between the imaging center and the edges (Fig. S2). Compared to higher electron dose rates (Fig. 1), the lower electron dose rate allows for more effective heat diffusion from the observation area. The electron dose versus time analysis of phase transition temperatures (Fig. 1f) shows that the phase transformation kinetics in 8CB liquid crystal depend on the electron dose rate, similar to the transition temperature shift with the temperature rate seen in conventional DSC measurements[23]. At an electron dose rate of 2.72 $e^-/nm^2s$ the phase changes were the most clearly visible among all the dose rate experiments and the N-I transition appeared at the shortest time, e.g. 32 s in Fig. 1f. This indicates that the time needed for the temperature increase to reach N-I transition temperature is increases with the dose rate value. On the other hand, the onset of the SmA-N transition (Fig. 1f) does not strongly depend on the dose rate value but to a certain extent on the irradiation time. The difference in SmA-N start time can be attributed to superheating at higher electron doses . The highest electron dose rate heats more quickly, so the nucleation of the new phase is shifted for higher temperatures. This results in a slight retardation of SmA-N occurrence in time with increasing electron dose rates.

This hypothesis is consistent with the observation that in uniform areas (Fig. 2) defects did not assist phase changes. The fact that the N-I transition begins in the center of the observation area (Fig. 2c) is linked to the geometry of heat dissipation in the sample[62]. It is known that when an electron beam transmits through a thin liquid sample, the electron dose distribution is uniform along the Z-axis (i.e., through the sample thickness)[64]. If we therefore consider temperature variations in this dimension to be negligible, as they are much smaller than the threshold for phase change, then the heat transfer takes place primarily in lateral directions (X-Y plane) to regions outside the irradiated area, implying that the temperature increase will be the highest in the center of the observation area[62]. This process will also be shown and discussed in the next part of this article.

### *Heat diffusion*

In order to better understand the thermal processes taking place in 8CB under the influence of an electron beam, we developed a simulation model corresponding to the observed situation, simplified to a 1D system. It should be noted that thermal simulations for liquid crystal systems do not yield consistent results, but only demonstrate the nature of thermal diffusion without fully accounting for the specific characteristics of liquid crystals. This is because the thermal parameters of planar and homeotropic alignments of liquid crystals depend on the spatial alignment. In our case, the thin liquid-crystal layer 8CB consists of a finite number of molecules forming a physicochemical system. The local addition of heat not only affects the temperature distribution of the entire system, but also influences the degree of packing and molecular reorientation. Hence, any numerical description using heat diffusion methods is limited, but we show below that it predicts sufficient temperature increase to lead to a phase transition.

In the simulation we described the 8CB liquid crystal using the parameters of specific heat $C_p$, thermal conductivity $k$, and density $\rho$ in accordance with Marinelli et al.[89] (see Table S1). The thin layer of 8CB exhibits homeotropic alignment; hence, in order to better represent the system, thermal parameters for the homeotropic alignment of the liquid crystal were adopted. Simulations of temperature distribution over time were calculated using the FDTD method based on the solution of the heat diffusion differential equation, with a localized heat source according to the equation:

$$\frac{\partial T}{\partial t} = \alpha \left(\frac{\partial^2 T}{\partial x^2}\right) + Q(x,t) \qquad (1)$$

where $\alpha = \frac{k}{\rho c_p}$ is the heat diffusion coefficient and Q(x,t) is the amount of heat transferred to the system in 1D model.

A spatially constant heat source value dependent on the electron dose rate was assumed in the central area of diameter 30 µm to match the experimental setup (c.f. Fig. 3a). For a value of 1 e-/nm²s electron dose rate, we discuss below that the equivalent amount of heat transferred to the experimental system is Q=4.46·10$^7$ W/m$^3$ [90]. The simulations were performed for a time step of 10 ms and a grid size of 1 µm, with a total time of 100 s. The temperature at the center point of the simulation was used as the main parameter to monitor the process, as shown in Fig. 3ab. We find that the temperature increase for successive electron dose rates is similar to that observed in the experiment and shown in Fig. 1 and Fig. 2. The simulations were performed in an isolated system with a total size of 5 mm. An increase in the dose rate results in a faster phase transition, but the nature of the changes obtained is non-linear in the initial period. The temperature profile shows a difference in detail compared to the experiment. In Fig. 3c, a time point of t = 27 s from the start of the simulation was selected at a dose rate of 2.72 e-/nm²s . At the point at which the N-I phase transition is observed, the nematic phase spreads to total length of 3 mm. This is a significant difference from the experimental data, although of course the experiment is not uniform over this large length. When the calculated system reaches the isotropic phase at the central point, it exhibits a nematic phase over a much larger area than was observed in the experiment. The increase in the area size for the SmA-N phase transition over time is shown in Fig. 3d, where an area increase due to heat diffusion can be observed. Despite obtaining the correct convergence of the numerical solution, we identified a problem consisting of the difficulty of simulating a system of tens of micrometers in size on a time scale of 100 s. According to the von Neumann stability condition $\Delta t \leq \frac{\Delta x^2}{\alpha}$, assuming the grid size to be $1 \cdot 10^{-6}$ m and the heat diffusion parameter value as $\sim 1 \cdot 10^{-7} \frac{m^2}{s}$, we obtain $\Delta t \leq \frac{(1 \cdot 10^{-6} \, m)^2}{1 \cdot 10^{-7} \frac{m^2}{s}} = 10 \cdot 10^{-6}$ s. A series of simulations were performed in a limited time of 1 s with a time step of 10 µs. The results obtained are in good agreement with the simulations performed for a time step of 10 ms for the first second of the simulation. The difficulties in simulating this system are evident. It is unclear whether the observed effects are related to the computational conditions or an inaccurate representation of the simulation conditions. Thermal parameters at the nanometer scale may differ and be affected by low pressure and near-field effects originating from the electron beam. The simulations lead to a plateau, while the differences in electron dose rate correspond well with experimental measurements.

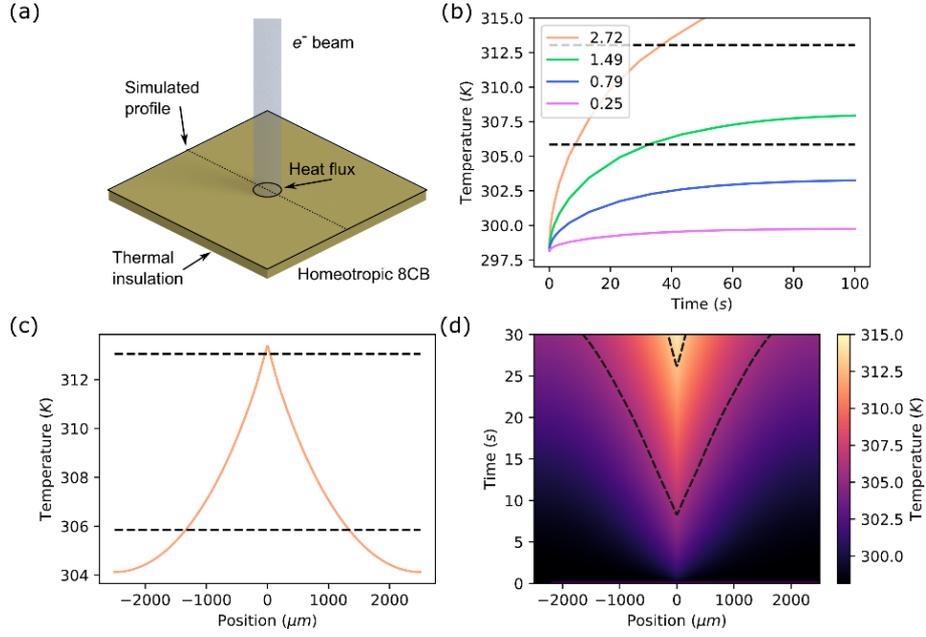

**Figure 3.** (a) Schematic representation of calculation parameters, the cylindrical electron beam generates heat in the sample center. (b) Temperature increase for the central point of the simulation at successive electron dose rate values. (c) Example temperature profile for time t = 27 s at a dose rate of 2.72 e-/nm²s. (d) Temperature distribution map over 30 s of simulation at a selected dose rate of 2.72 e-/nm²s. The SmA-N and N-I phase transitions are marked with dashed lines.

While the stopping power (S) for 200 keV electrons is nearly identical for both water (2.80 MeV·cm²/g) and 8CB (2.78 MeV·cm²/g), their thermal properties differ substantially (Table S1). The maximum temperature rise ($\Delta T_{max}$) in a material subjected to electron beam irradiation[63] can be determined by the material's stopping power, specific heat capacity, and thermal diffusivity, as described by the following relationship:

$$\Delta T_{max} = \frac{SI10^2}{\pi \alpha C_p}\left(\frac{1}{4} + \frac{1}{2}Ln\left[\frac{L}{a}\right]\right) \qquad (2)$$

where S is the density-normalized stopping power [MeV·cm²/g], I is the current [C/s], a is the radius of the beam [m], $\alpha$ is the thermal diffusivity [m²/s], $C_p$ is the specific heat [J/kg·K] and L is the radius of the disk of the outer boundary of the computational domain (for more details check the source reference [63]).

The specific heat capacity of 8CB is approximately half that of water (4189.9 J/kg·K), meaning it requires significantly less energy to heat. Furthermore, the thermal diffusivity $\alpha$ of 8CB is highly anisotropic and temperature-dependent, as presented in Table S1. At room temperature, its value (~0.0012 cm²/s) is close to that of water (0.00143 cm²/s). However, during heating and phase transitions, the thermal diffusivity of 8CB can drop to 0.0006 cm²/s or even lower. This combination of a low specific heat capacity and a variable, often low, thermal diffusivity leads to a maximum temperature rise in 8CB that can be 2.5 to 5 times higher than in water in the same imaging conditions. However, the mechanism by which energy is delivered to 8CB and water is very likely significantly different, as evidenced by the poor agreement between the simulation and the experiment, particularly regarding heat distribution within the sample. This discrepancy between simulation and experiment allows us to hypothesize that either the thermal properties of 8CB at the sub-microscale differ significantly from their macroscopic properties, or the material is capable of absorbing more energy from the electron beam than it is estimated using stopping power, or both. There is also no doubt that the mechanism of pulsed energy delivery by STEM itself may be difficult to represent as a continuous heat source. This topic undoubtedly deserves further exploration.

*In situ laser-induced heating*

The relatively high temperature increase we require to explain the phase transitions in 8CB LC (with a maximum of about 15°C for the N-I transition) during these LP-STEM experiments might raise some concerns[64]. However, we have assumed in the calculations that any effects are due only to sample heating from the electron-beam irradiation, without considering radiolytic effects. According to the literature, for bulk 8CB tested in a 7000 keV linear electron accelerator, high electron doses (in the order of 6 kGy and more) cause the material to undergo partial radiolysis, inducing detachment of CN groups from some 8CB molecules (8CB$^+$) by chemical bond cracking[91]. Such effects were found to affect the thermal properties of the mesophases, phase transition temperatures, electrooptic and dielectric anisotropies and molecular relaxation frequency depending on the used dose conditions[66].

It is therefore important to compare the phenomena we have observed under electron beam irradiation, which may also involve irreversible radiolysis effects, with transformations driven only by heating. To accomplish this, we therefore developed a new imaging approach. We performed control experiments with a laser, working at 445 nm wavelength, as an additional heating source during STEM imaging. The procedure is described in Methods and Fig. S6. It is important to emphasize that in our system, the entire sample was illuminated, not just the chosen (S)TEM imaging area. Generally, pure 8CB does not absorb 445 nm wavelength light[91]. Therefore, it can be assumed that the overall temperature increase of the LC was induced by heating of the sample surrounding substrates, i.e., amorphous carbon layers and Cu grids. This approach provides more uniform heating across the entire system, without accounting for local effects arising from direct light absorption by the LC. We note that polarized laser light is known to induce the reorientation of LC molecules in the nematic phase via third order nonlinear optical effect called optical Freedericksz transition (OFT)[92,93]. Observing this effect requires much higher light powers than those used in this experiment. Additionally, the substantial light scattering and reflection inside the microscope column significantly reduce the initial laser beam's polarization.

We captured and analyzed LP-STEM images with the laser switched on and off. We found that the laser highly accelerated the phase transition sequence. This made it difficult to record and compare specific phase changes so we performed the experiments in a series of steps.

First, the electron beam induced the standard SmA-N and N-I phase transition sequence at the dose rate of 2.72 e$^-$/nm$^2$s (Fig. 4a). Then, the laser light was switched on once the isotropic phase appeared (Fig. 4b). The experiment was performed this way, because the laser highly accelerated the phase transition sequence, making it impossible to record and compare specific phase changes. The laser illumination caused a vast increase in I phase volume in the observation area (of about 70 %, Fig. 5be), which occurred in a time shorter than one full STEM image frame in the series (i.e. 0.84 s, see supplementary video 5[82]). With further light illumination for 15 seconds, the I phase covered almost 100% of the observation area. Switching off the laser light resulted in an instant decrease in temperature and a shrinkage of the I phase by nearly 60% (Fig. 4ce). The structure of the N and I phase mixture was inhomogeneous and composed of I phase and small droplets of N phase inside phase I, meaning that I-N transformation happened quite uniformly within the sample. This feature is a result of a sudden temperature drop, contrasting with N-I transformation observed in the previous experiments with slow heating. With further electron beam irradiation, the I phase then propagated slowly, the N mesophase droplets coalesced and then slowly disappeared (Fig.4d). The process was repeatable when turning on and off the laser light, with an overall increase in I phase content over time (see supplementary video 5[82]).

These experiments confirmed that the growth of the I phase can be induced by both the electron beam and the laser reversibly, and the observed processes were a result of temperature increase to a predominant extent. When performing STEM imaging, we cannot eliminate the influence of the electron beam to track a process that is entirely dependent on another heat source. However, we were able to visualize the effect of minimizing the electron dose on reversibility of the phase transitions. This will be described in the next section.

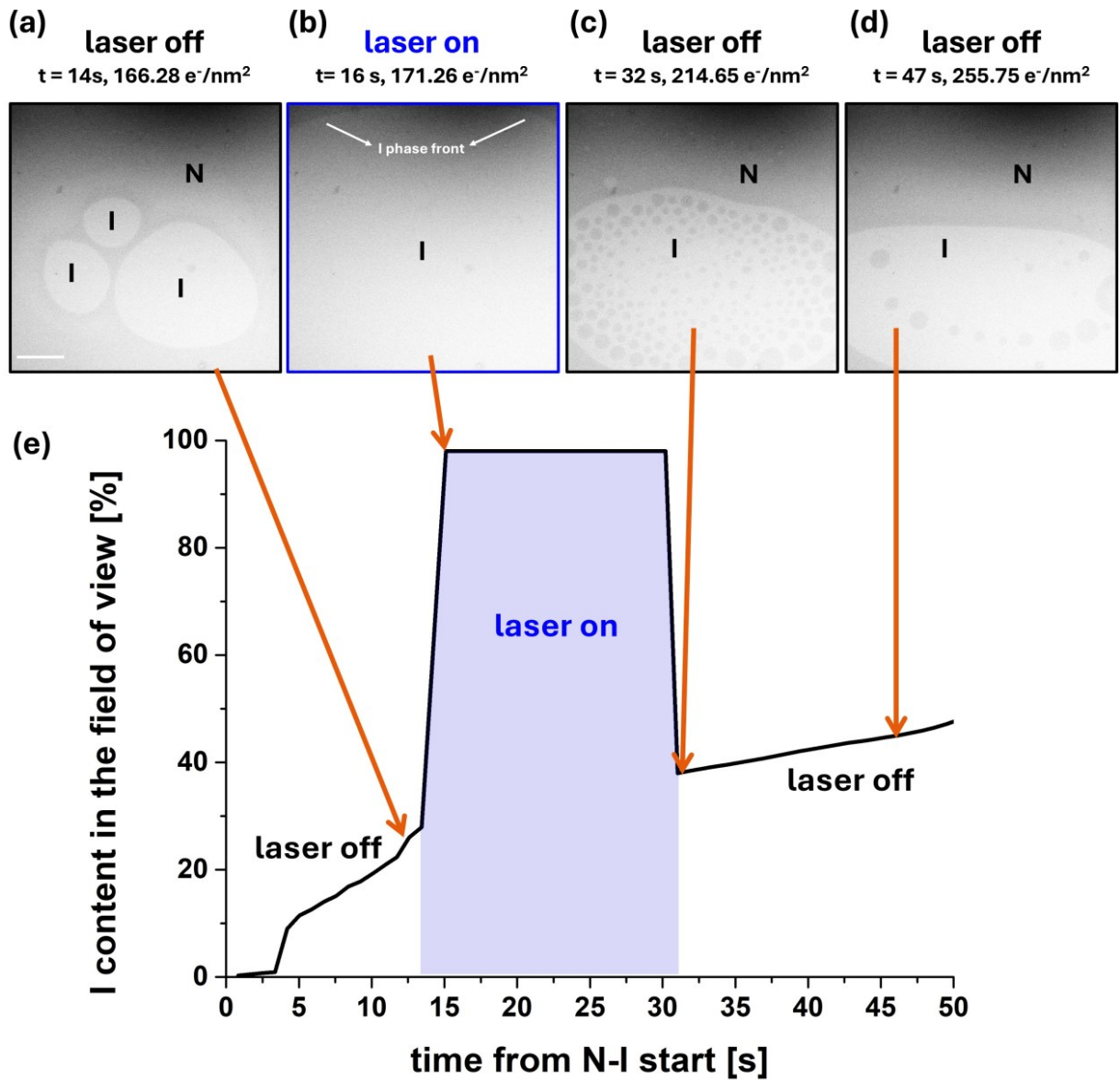

**Figure.4.** LP-STEM of N-I phase change accelerated by laser heating. The electron dose was constant with at 2.72 e⁻/nm²s. The time shown in the images and on the plot was counted from the start of N-I; the electron dose is the cumulative electron dose, including the pre-imaging time. (a) The growth of the I phase under electron beam influence only. (b) Sudden increase in I phase volume that emerged upon laser illumination. (c) Immediately just after the laser was switched off. (d) The state upon further electron irradiation when, the I phase became homogenous again. (e) The quantitative analysis of I

phase percentage in the observation area in time with laser switched on and off. For the full-time fast series, see Supplementary video 5[82].

*Reversibility of phase transformations*

As the I-N phase transformation was observed upon switching off the laser, an additional experiment was conducted to assess the reversibility of the process. Single STEM images were taken during the cooling the sample after switching off the electron beam. The same area was imaged with and without constant additional laser heating at a wavelength of 445 nm. The reversibility was measured by estimating the I phase content in the observation area [%] and its decrease in time. The images were taken with a relatively low dose of 2.28 e$^-$/nm$^2$ per image to minimize the effect of the single scan on the reversibility process. The raw images were edited for enhanced visibility of changes by subtracting a reference image of the 100% N phase. The resulting processed images are presented in the first rows of Fig. 5ab. Their false-colored versions are also displayed in the bottom rows for easier brightness change interpretation, where warmer colors indicate higher brightness in the grey images. The raw data is provided in Fig. S5 in the Supplementary Information.

In case of no additional heating, the greatest decrease in I phase content (by about 90%, see Fig. 5ac) in the first 90 s of cooling (equivalent to beam-off time in Fig. 5). With further cooling the I phase shrinks circularly and almost disappears completely after 900 s (Fig. 5a). Different results were obtained in case of simultaneous laser irradiation. The largest decrease (about 40 %) was still observed within the first 90 s after switching off the electron beam (Fig. 5bc). After that time, the laser illumination was long enough (time equivalent to beam-off time) to induce further growth of the I phase despite the electron beam's absence. An interesting contrast feature is observed with longer laser irradiation times, where the differences between the N and I phases become less distinguishable. With extended laser irradiation in the absence of an electron beam, the overall brightness of the images increases (Fig. 5b), leading to an almost invisible phase boundary between the N and I phases as the sample approaches the transition temperature. This is probably a result of constant, relatively slow heating of the whole sample. The brightening of the N phase can be related to the effects described in Fig. 2, where the intensity profile indicates a vast increase in brightness right before the start of the N-I transition. In the case of localized electron irradiation only (Fig. 2b), the effect is mostly visible in the center, while in case of homogenous laser heating (Fig. 5b) the whole N phase surrounding I phase becomes brighter. It is also possible that minor thickness changes take place due LC movements induced by a temperature rise.

We also note that before laser illumination an image of the region surrounding the area scanned in Fig. 5a, taken at lower magnification (Fig. 5d), revealed some N phase near around the scanned area and unaffected SmA phase at further distances. Such observation suggests that the heat dissipates strongly from the imaging area, leading to I phase formation mainly in the center.

The results of the experiments with laser heating show that the N-I phase change is fully reversible once the heating sources are removed, leading to a few conclusions. First, the observed processes are thermally induced, and the heat caused by the electron beam scanning the field of view dissipates in the X and Y axes out of the observation area. This leads to a higher temperature increase in the center, where the N-I phase change appears first. In case of additional heating with a laser, the process of N-I is accelerated significantly. The presence of a two-phase mixture, with oval areas of the N phase suspended within the I phase, indicates that rapid cooling upon turning off the laser alters the appearance of the I-N transition.

According to the literature[91], an electron beam with a dose on the order of magnitude of 6 kGy or more can induce the detachment of the CN group, leading to the formation of a cation 8CB$^+$. In our experiment, the electron dose of 199 e-/nm² at the N-I change shown in Fig.5a for 0s beam-off time corresponds to 14 MGy (calculated for the accelerating voltage of 200kV [94], a stopping power[90] of 2.78 MeV·cm²/g and thus the correction factor of 4.45), which is a value about three orders of magnitude higher. This suggests that radiolysis does happen in the LC sample, leading to the formation of cyanide anions. However, in the absence of oxygen in the sample, the ions do not undergo further oxidation that would cause an irreversible LC degradation. Therefore, an equilibrium between ionized and primary 8CB molecules is established during constant electron irradiation. While these molecular changes have been shown to cause a small shift in phase transition temperatures[91], the overall reversible behavior observed in our preliminary studies of 8CB suggests that widespread, irreversible LC degradation is not the dominant mechanism under our controlled low-dose conditions. Therefore, we conclude that the radiolytic damage is not dominant in the observed phenomena, as no permanent changes appear in the sample. Nevertheless, this conclusion needs more study accompanied by, for example, infrared or Raman spectroscopy.

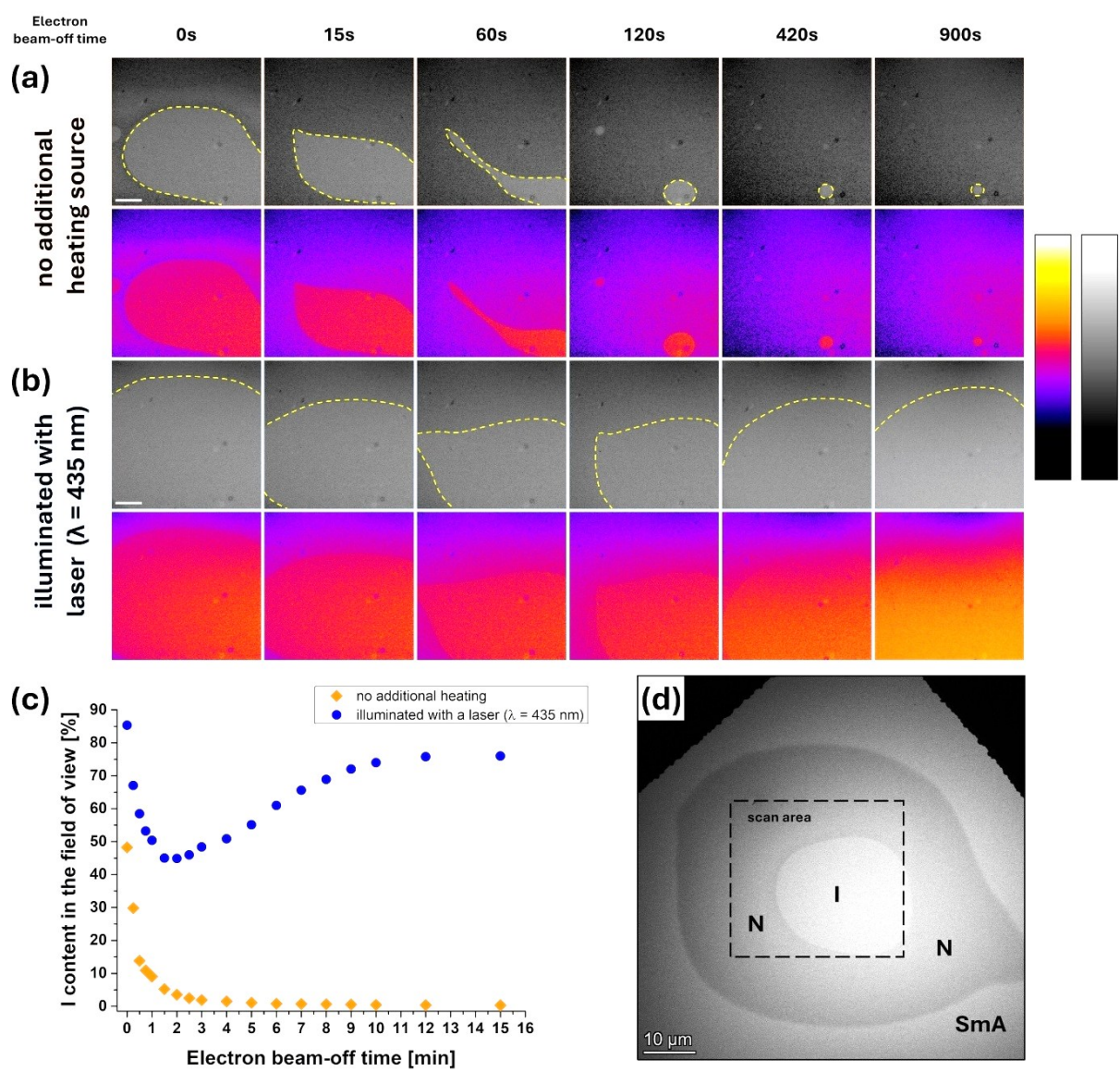

**Figure 5**. LP-STEM of N-I phase change reversibility. In (a) and (b), the upper rows show images with a subtracted background of 100% N phase (see Fig. S5 for raw data), and the lower rows show the same images with a false color filter for easier distinction of contrast changes in time with and without laser. A few impurities cause artifacts after subtracting due to sample bulging (dark and light points next to each other, indicated in Fig. S5). (a) Reversibility over time without any additional heating source – the sample was cooling down as the electron beam was switched off. At 0s, the typical bright ring (described in Fig. 3) around I mesophase can be seen. The I phase was gradually shrinking over time. (b) I phase evolution over time under a constant laser irradiation. A slight decrease is observed in the first 120 s after switching off the e-beam; the I phase increases due to laser heating. (c) Dynamics of I phase area percentage changes with time after switching off e-beam irradiation (orange color squares) and under auxiliary laser irradiation ($\lambda$= 445 nm at 10$\mu$W power, blue color circles in the plot) of the studied area. (d) The lower magnification STEM image showing the outside region of the scanned area (30x30 $\mu m^2$) in the experiment. N phase surrounds the scan area, suggesting the heat dissipation into that region. The scale bar of 5 $\mu$m is the same for all images in each row in (a) and (b).

### *On contrast origin of the N phase*

An interesting feature observed in our LP-STEM experiments is the dissimilarity in the electron-scattering process across the different phases of 8CB liquid crystal. Intuitively, SmA phase should give the highest contrast in STEM as this phase is the most organized and closest to crystal. However, in the case of a very thin LC layer other subtle factors could affect the observed contrast as well. While SmA and I phases give similar, weak contrast (expected for organic LC sample) based mainly on the thickness variations, a different appearance was observed for N. The N phase gives the highest contrast (strongest scattering and darker appearance in observed HAADF signal) compared to the other two phases, even though the overall density of LC in this phase is lower than for SmA and higher than for I. An additional experiment was performed to compare electron scattering by both phases, observing the diffraction pattern in STEM in two different phases. For a location of the same thickness and exposure, the nematic phase exhibited a lower central reflection intensity (Fig. 6), but with a higher number of electrons scattered at small angles. In contrast, the isotropic phase exhibited a higher central beam intensity and fewer scattered electrons. This is consistent with the observed contrast characteristics of both phases in STEM.

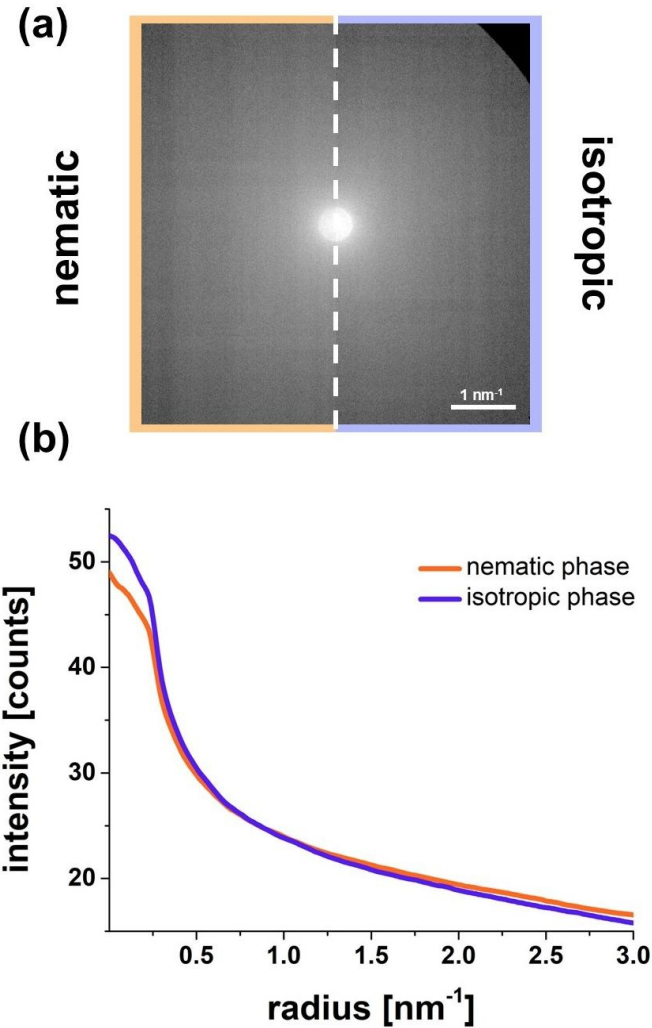

**Figure 6.** The comparison of electron scattering in N and I phases of 8CB LC. (a) The central diffraction spot in STEM mode has lower intensity in N phase, suggesting the higher scattering compared to I phase. (b) Mean radial scattering profile of N and I phases. N phase has notably lower intensity in the central diffraction disk than I phase, which slowly changes with the radius increase. This indicates that in N phase, transmitted electrons are more likely to be scattered more at higher detection angles.

The interpretation of several factors that are characteristic for liquid crystal mesophases can be discussed to rationalize such a non-intuitive contrast change. Nevertheless, mesophases are described by an orientational order parameter (S) fluctuation. The value of S gradually decreases from 1 in the solid state crystal. It steeply decreases with temperature increase, as the material transitions from SmA to N phase, and then abruptly reaches S = 0 in the isotropic phase. SmA phase is generally more ordered than nematic one as the LC rod-like molecules follow the director $\hat{n}$ direction and exhibit a positional order forming a layered structure. The molecules have more freedom in the N phase as they are only aligned along the $\hat{n}$ and lack positional order. The absence of long-range positional order grants nematic molecules the higher freedom to rotate and translate. The density of 8CB varies with phase and temperature. It is approximately 1.00 g/cm³ near room temperature and is slightly lower (around 0.98 g/cm³) just below the nematic-isotropic transition temperature (see Table S1 for plot). Keeping this in mind, we can conclude that the molecules in N phase occupy more space, and they could reorient[95] along the electric field produced by the electron beam[96] due to large positive dielectric anisotropy $\Delta\varepsilon \approx 9$. The strength of the electric field induced by the STEM probe decreases with increasing distance from the beam and its direction points mainly along the sample plane due to Auger and secondary electrons emissions from the sample surface[96]. Therefore, in the presence of e-beam electric field, we propose that the LC molecules align planarly, i.e., in the plane of the sample. In case of such subtle contrast variations in low-density LC sample, we suggest that some stronger inelastic electron scattering on planarly oriented 8CB molecules may contribute to the darker contrast appearance for the N phase, similarly to some polymers containing aromatic systems observed in energy-filtered TEM[97–99]. Additionally, this less organized molecule distribution and librational motions could result in slight thickness changes on the molecular level. There is no doubt, however, that this topic requires more detailed analysis.

**Conclusions**

In summary, the novelty of this work lies in the detailed in-situ imaging of electron beam-induced SmA-N and N-I phase changes dynamics, together with the defect formation and accompanying surface phenomena. The hypothesis of a distinctive N phase contrast origin is discussed, emphasizing N phase reorientation in the beam's electric field and electron interactions with planarly oriented biphenyl rings of 8CB molecules, which causes higher electron scattering of this phase. By developing a novel laser-illumination system in TEM, we show the observed phase transitions are predominantly driven by temperature increase, as they occurred during both electron beam irradiation and with the laser as an additional heating source. The calculations show that the nanoconfined 8CB differs in thermal properties significantly compared to the bulk. The phase changes were fully reversible once the heating source was removed, which suggests not only the temperature-dependent nature of the phenomena, but also excludes irreversible electron beam damage during the imaging. However, the radiolytic effects should be studied in more detail, as ion formation could affect sample heating.

Our work lays a foundation for further in-situ LP-(S)TEM studies of liquid-crystalline materials, including their advanced nanocomposites. TEM can be used as a powerful device for studying LC response to electric and magnetic fields, other phase transitions dynamics, nanomaterials assembly in the LC matrix, or their general interaction with electrons for fundamental knowledge. The application of laser-induced heating in the TEM offers a pathway for exploring reaction mechanisms and radiolysis effects. We also see promising prospects for the possibility of controlling LC phase transitions with electron beam at the sub-micro and nano scale, providing an opportunity to design optical systems with unique properties.

**Methods**

<u>Transmission Electron Microscopy</u>

A Talos F200i scanning transmission electron microscope (ThermoFisher Scientific) operated at 200 kV was used for liquid-cell STEM (LC-STEM) studies. The device is equipped with a borosilicate window and a mirror within the pole piece, which allows laser light to be directed to the sample plane during observation (Fig. S6). In the experiment with additional sample heating with laser light, a CNI MDL-III-445 continuous-wave laser with a maximum power of 1 W and a wavelength of 445 nm was used. The light intensity in the sample plane in the described experiment was approximately $4 \cdot 10^6$ W/m².

The liquid-cell samples were prepared using carbon encapsulation with copper grids covered with 25 nm amorphous carbon (Agar Scientific)[74]. First, the grids were glow-discharged for 20 s at 30 mA plasma current. The application of LC on a TEM grid can be performed in two ways. Either 0.5 μl of the solution of pure 8CB dissolved in chloroform (1 %wt.) can be dropped on the grid and left for evaporation at 45°C or a minimal amount of LC can be placed directly using a needle tip. The first method provides a much thinner film, but in that case, LC forms inhomogeneous "islands" that spread quickly upon electron irradiation (see Fig.S7 in Supplementary Information). The needle preparation could result in a much thicker sample, but provides more homogeneous observation areas. In our experiments, we used both methods, and the final observation areas were selected over a few samples depending on the enclosure's effectiveness. After application of LC on the first grid, the second, clean one was placed on top to form a "sandwich" and held vertically between two cross tweezers[74] at 45 °C for 5 min to obtain a more homogenous layer. After that, the sample was put for 1 min into the freezer set at -12 °C. The imaging was performed on a sample kept at room temperature for 5 min after being taken out of the freezer.

The observations were performed in "low magnification" STEM mode, in which the objective lens power is lowered. The magnetic field in the sample area is minimized to 150 mT to avoid possible disturbances to the observations in the liquid-crystalline state. The camera length was the shortest possible, namely 4.7 m, and the collection angle of annular detector was 1-8 mrad. The central diffracted disk was wider than the inner diameter of the high-angle annular dark field (HAADF) detector, so the imaging was performed in quasi-annular bright field STEM mode. The phases that scatter electrons more appear darker than less scattering phases. This is because the more scattering phase deflects a higher proportion of the central, unscattered beam away from the HAADF collection angle area, causing a reduction in intensity of central spot. This beam and detector arrangement proved the most advantageous for imaging minor differences in contrast (scattering) between individual 8CB phases. The electron dose for STEM images was calculated from the beam current and dwell time for each frame[94]. For data shown in Fig. 1ce, Fig. 2, and Fig. 4, the images were acquired as fast series at 2048x2048 resolution, with a dwell time of 200 ns and a frame time of 0.84 s. For a detailed electron dose calculation, the acquisition of specific areas was started at very low magnification, and the total electron dose included the dose used for pre-imaging. The supplementary videos show raw data captured using a fast series and are available at the data repository[82]. The frame rate (fps) for videos 1–4 is 10, while video 5 has a frame rate of 25. The supplementary video for Fig. 1d (the lowest electron dose rate of 0.25 $e^-/nm^2s$) is not provided because the process was slow, and the individual images were taken instead of a series. Image analysis was performed using Velox (ThermoFisher Scientific) and Fiji software[100]. All images were edited with an averaging filter with 4x4-pixel radius. In terms of time-resolved series, the contrast and brightness remained fixed through the experiment.

The experiments exploring the origin of LC phase contrast and electron scattering were made in the same imaging mode as previously described. In the desired field of view we performed imaging until achieving wide region of N phase. The beam was then immediately blanked, moved to the location of the N phase, and one exposure of STEM diffraction pattern was achieved. The central beam appeared as diffused disc because of the convergent STEM electron beam. Then imaging continued until the I phase region appeared in the same area. The beam was then blanked again, moved tens of nanometers away from the previous position, and STEM diffraction was recorded one more time. The beam position between two diffraction patterns was shifted slightly to ensure the same impact of any accumulated layer of contamination products for two images taken adjacent to each other. STEM scans were performed before, between, and after the diffraction to ensure that the specific phase was present at the analyzed area based on contrast at the beam location. The exposure time and detector resolution and binning were adjusted to achieve a reasonable signal level and a stable phase composition at the beam stop.

## Acknowledgements

The research was funded by the Ministry of Science and Higher Education (Poland) grant PN/01/0121/2022 from the program Perly Nauki. FMR and AŻ acknowledge the MIT-Poland Lockheed Martin Seed Fund, Polish-American Fulbright Commission and Institute of International Education (Fulbright Senior Award). OK thanks Amelia Maj and Patryk Obstarczyk (Wroclaw University of Science and Technology) for their assistance with the laser experiments, Anna Kotowska for help with figure 1 and Lunek Ekri for a valuable discussion. AŻ would like to thank Malwina Sikora (Nanores, Wroclaw, Poland) for lamella preparation.

**SUPPLEMENTARY INFORMATION**

**In-situ liquid-phase scanning transmission electron microscopy of 8CB liquid crystal beam-induced phase transformations**

Olga Kaczmarczyk[1], Konrad Cyprych[2], Dominika Benkowska-Biernacka[1], Rafał Kowalczyk[3], Katarzyna Matczyszyn[1], Hanglong Wu[4], Frances M.Ross[4], Andrzej Miniewicz[1], Andrzej Żak[14]

[1] Institute of Advanced Materials, Wroclaw University of Science and Technology, Wroclaw, Poland

[2] Soft Matter Optics Group, Faculty of Chemistry, Wroclaw University of Science and Technology

[3] Department of Bioorganic Chemistry, Faculty of Chemistry, Wroclaw University of Science and Technology

[4] Department of Material Science and Engineering, Massachusetts Institute of Technology, Cambridge, Massachusetts 02139, United States

*Polarizing microscopy*

Images of 8CB enclosed between two amorphous carbon films (Fig. S1) were captured using Olympus BX60 polarizing microscope equipped with a temperature-regulated Linkam LTS120 stage.

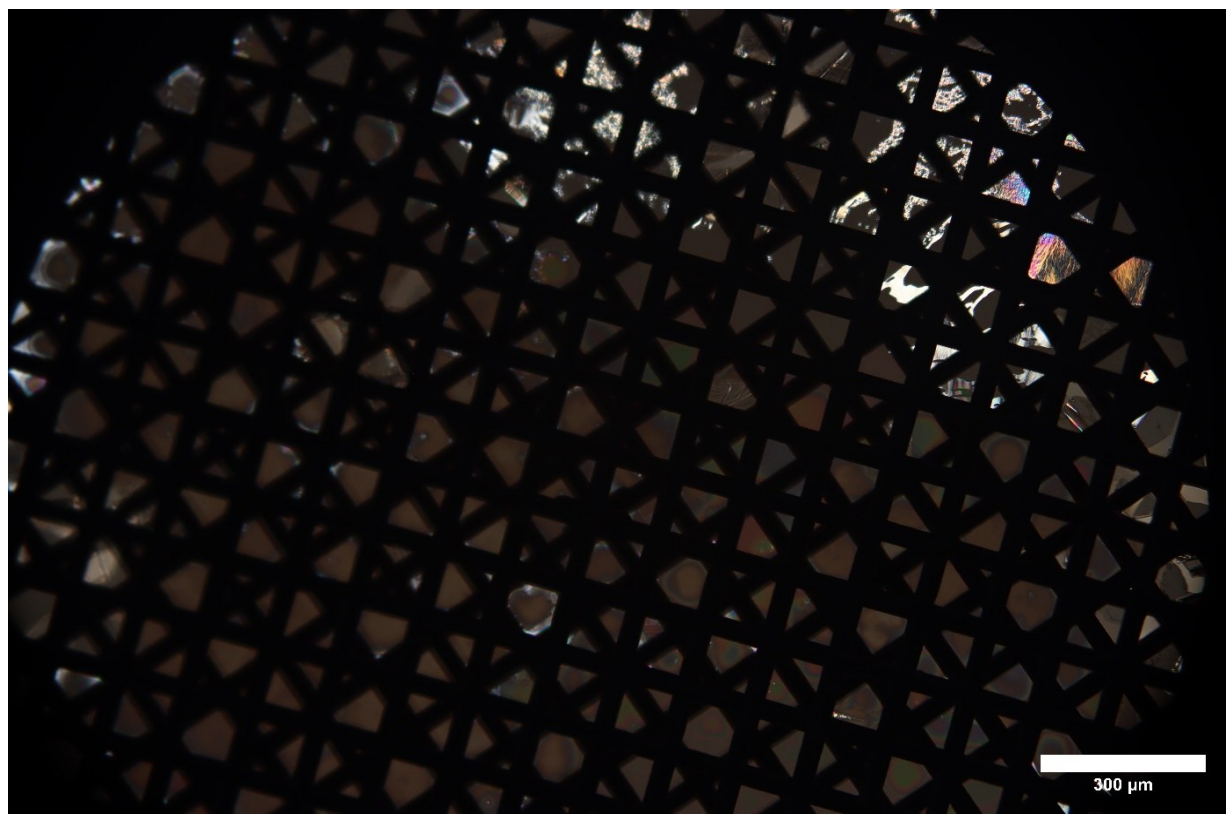

**Figure. S1** Polarizing microscope images of the 8CB sample sandwiched between two TEM grids covered with amorphous carbon. The image was captured with the polarizers crossed. The thin, electron transparent areas are not visible on the light microscope image due to the low thickness. The temperature was 25°C implying that the material was in the SmA phase.

*Areas for defect formation and phase transition observations*

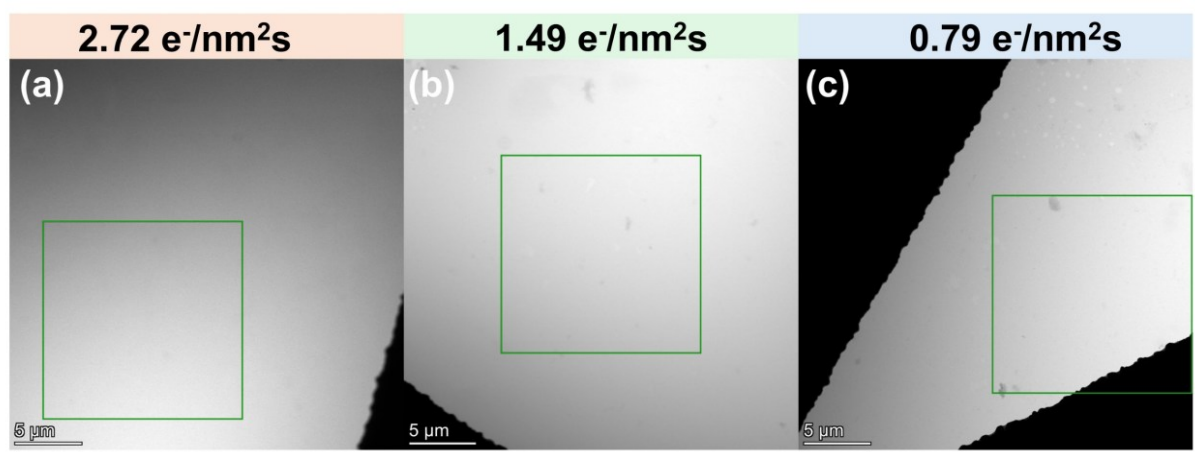

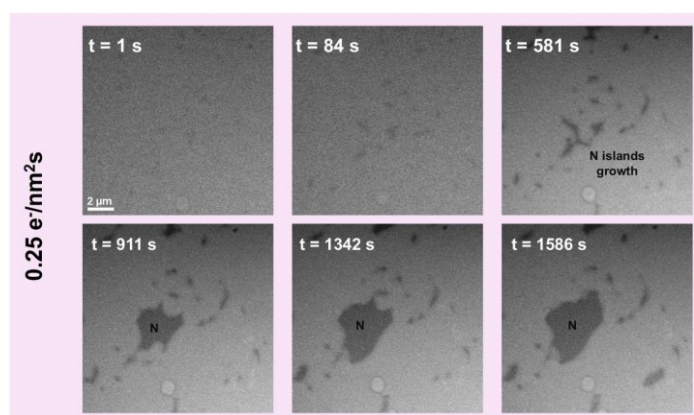

**Figure S2.** The crop areas for Fig.1 in the main article data for electron dose rates: (a) 2.72, (b) 1.49, (c) 0.79 e-/nm2. The observation area was 29.7 μm x 29.7 μm and the cropped areas were 15 μm x 15 μm. The black areas are Cu grid. (d) Data for Fig.1(f) in the main article, corresponding to the purple plot. No full phase change was observed during whole 25 min of observation.

*Sample thickness estimation*

The sample thickness was measured after drop casting gold nanoparticle solution on both outer sides of the carbon/LC/carbon sandwich liquid cell. After the sample dried, micrographs were taken at sample tilts of approximately -20°, 0°, and 20° for the areas where the phase transition could be observed, and the parallax method was used to calculate the thickness. The images were superimposed in pairs of -20°/0° (fig. S3a) and 0°/20° (fig. S3b). For each pair, families of particles belonging to the upper and lower surfaces were observed. Knowing the difference in distance between the nanoparticle families located on both sides and the rotation angle, the local distance between the surfaces (and therefore the total sample thickness in the field of view) was determined to be 85 ± 27 nm . However, this thickness additionally includes the thicknesses of the two carbon films.

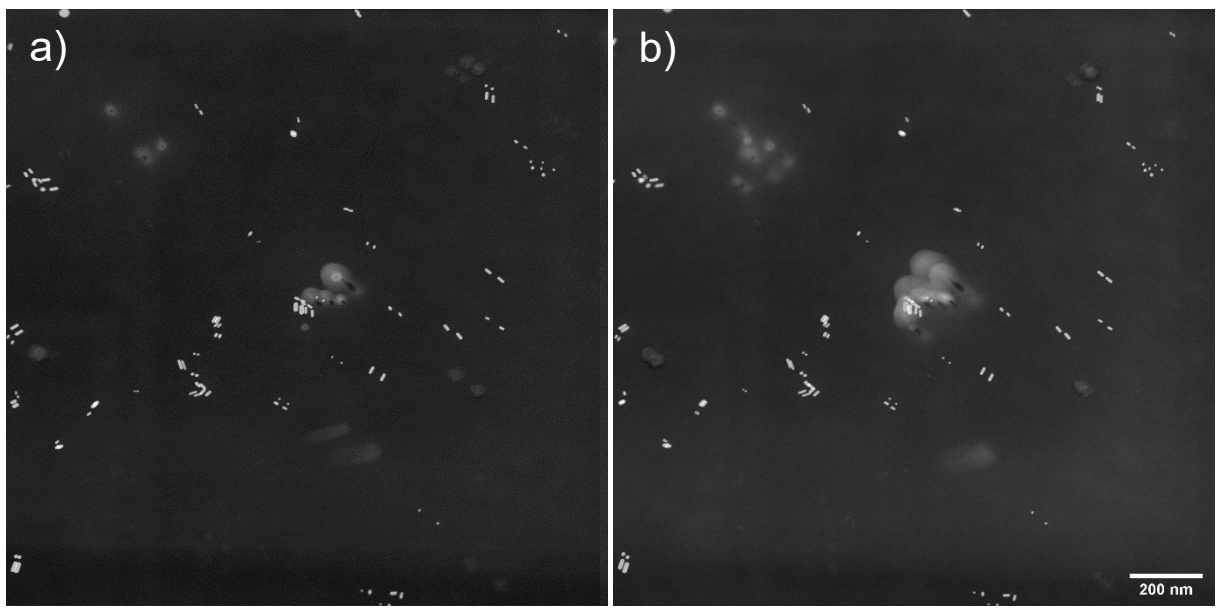

**Figure S3.** Superimposed micrographs with tilts of -20°/0° (a) and 0/20° (b).

According to the manufacturer, Agar S160 support grids have an amorphous carbon film approximately 28 to 30 nm thick. To be certain, one clean grid was sputtered with gold on both sides using magnetron sputtering and then cross-sectioned using a focused ion beam (FIB). The cross-section revealed that the region thinned most by FIB exhibited an artifact in the form of a thickened carbon layer, different in contrast from the base material (Fig. S4a). Observations of the region less thinned by FIB at the edge of the lamella confirmed that the carbon film was approximately 25 nm thick (Fig. S4b). Considering that the carbon grids were subjected to gentle plasma cleaning prior to preparation (which, according to previous estimates, thins the substrate by approximately 20%), we estimate that the thickness of the liquid crystal material at the previously measured location ranges from 20 to 40 nm and may be locally variable, or may even change due to the phase transition of the liquid crystal material. It is also possible that in some of the observed locations the sample thickness deviated from our estimate.

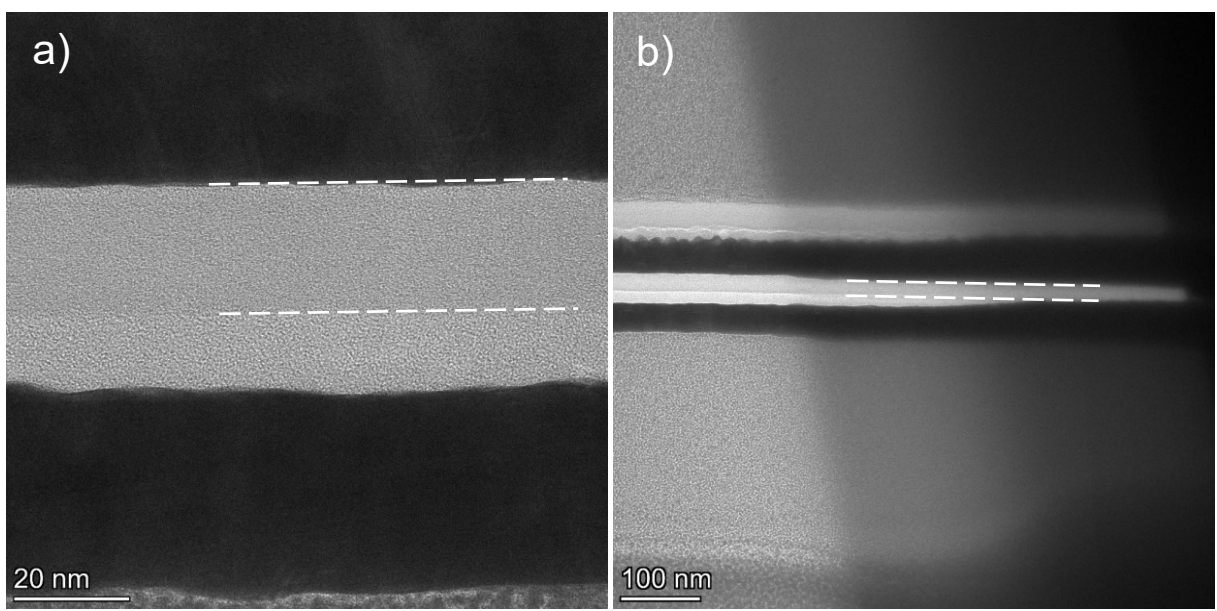

**Figure S4.** FIB cross-section of carbon film sputtered with gold. (a) Fully thinned region of the lamella. (b) The edge of the lamella that is less thinned. In both regions, the true carbon thickness is marked.

*Electron dose calculation*

The electron dose D was calculated for STEM images basing on the equation:

$$D = \frac{I\tau}{ed^2} \quad (1)$$

Where *I* is probe current, *τ* is dwell time, *e* is elementary charge, and *d* is pixel size. The images were recorded at 200 ns dwell time and 2048x2048 resolution.

Table S1. Comparison of key physical and chemical properties of water and 8CB LC, which can influence the sample heating during electron imaging.

| Property | Water | 8CB (4'-octyl-4-cyanobiphenyl) | references |
|---|---|---|---|
| Molecular structure | 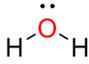 | 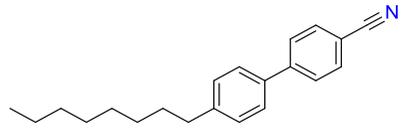 | 1,2 |
| Molecular formula | $H_2O$ | $C_{21}H_{25}N$ | 1,2 |
| Valence electrons | 6 | 88 | 1,2 |
| π electrons | *none* | 12 delocalized π electrons | 1,2 |
| Phase transitions | Solid ↔ Liquid (0°C) <br> Liquid ↔ Gas (100°C) | Crystal ↔ Smectic A (SmA) (21.3°C) <br> Smectic A (SmA) ↔ Nematic(N) (32.7°C) <br> Nematic(N) ↔ Isotropic liquid(I) (39.9°C) | 1,3 |
| **Density** 25.0 - 40.0 °C [g/cm³] | **0.997 – 0.992** | *Changes with temperature for each phase* **SmA: 0.998-0.985; N: 0.985-0.976; I: 0.976-0.965** 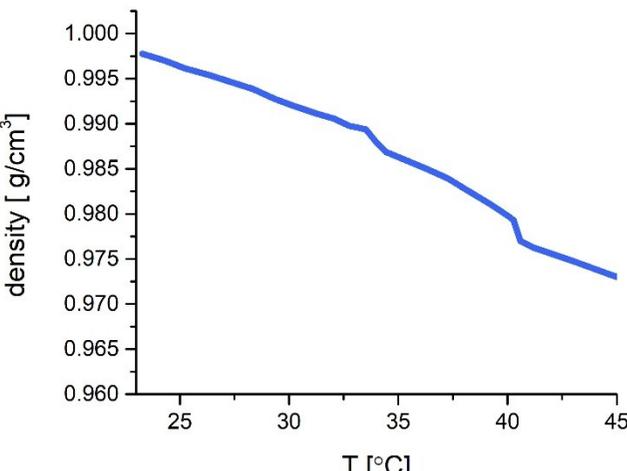 | 1,4 |
| **Specific Heat Capacity [J/g·K]** | *The value is relatively constant in the temperature range* **4.184** | *Specific for mesophases with peaks in phase transition temperatures* 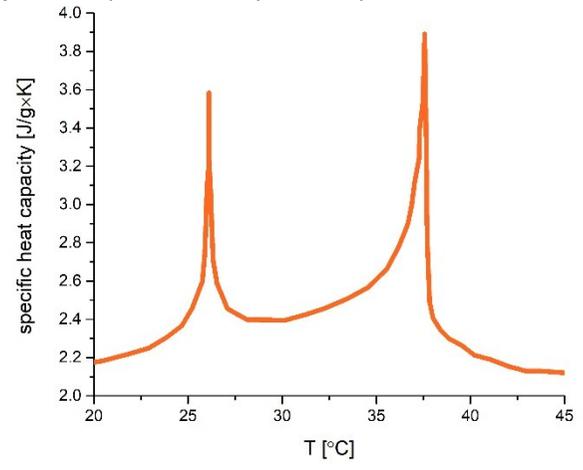 | 3,5 |

| Thermal conductivity 25.0 - 40.0 °C [W/m·K] | 0.607-0.669 | **Anisotropic**, the values slightly increase with temperature<br>**SmA and N**:<br>**0.12-0.14** *planar orientation*<br>**0.25-0.20** *homeotropic orientation*<br>**I: 0.15**<br>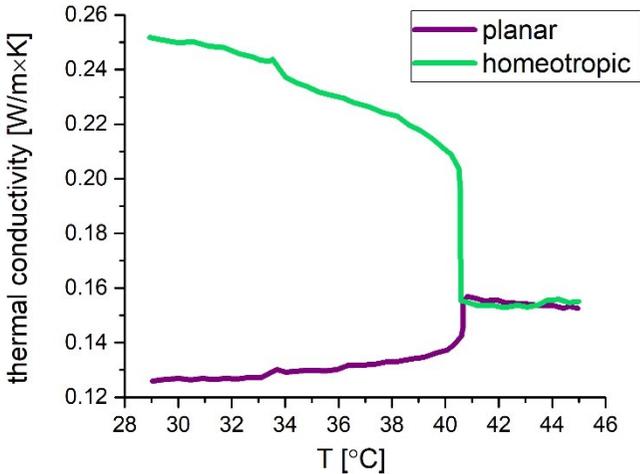 | 1,6 |
|---|---|---|---|

| **Thermal diffusivity** 25.0 - 40.0 °C [cm²/s] | *The value is relatively constant in the temperature range* **0.0014** | *Anisotropic* Values smaller than for water, in the range **0.0004-0.0010** 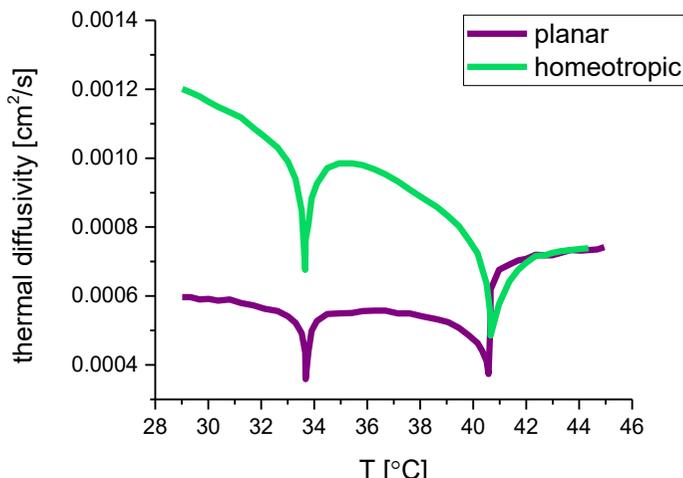 | 6,7 |
|---|---|---|---|
| **Electrical conductivity** [µS/cm] | 0.055 – 0.100 (ultra-pure water) | *The values slightly increase with temperature* **SmA: $5.6·10^{-5} – 2.5·10^{-4}$** **N: $3.4·10^{-4} – 6.8·10^{-4}$** *For frequency dependence: check the reference* | 8,9 |

*The plots presenting the density and thermal properties of 8CB LC are reconstructed from literature* [3,6,10].

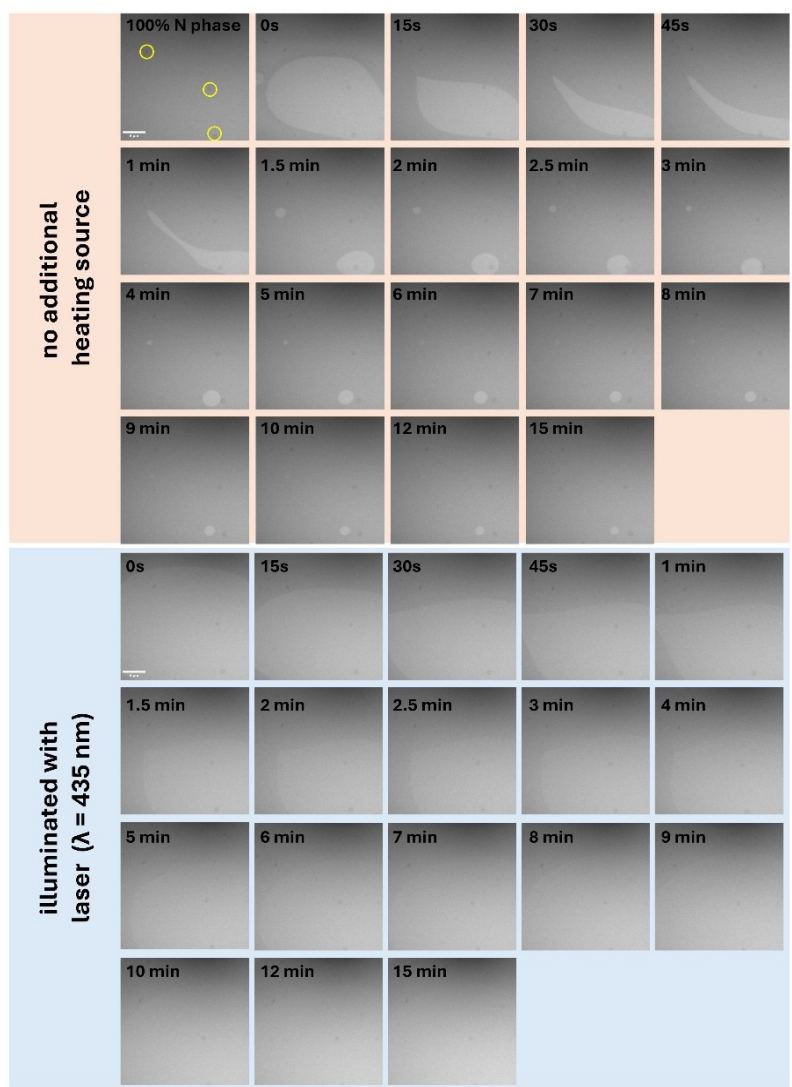

**Figure S5.** Raw data for the plot and edited images of Fig.5 and an additional image of 100% N, which was subtracted from the images to obtain the differences in contrast with the electron beam-off time. All images were edited with an averaging filter with a 4 pixel radius using Fiji software. No further editing was performed. Small impurities that change position with time (inside or outside the liquid cell) causing minor artefacts in the subtracted images in Fig.4 are marked in 100%N image.

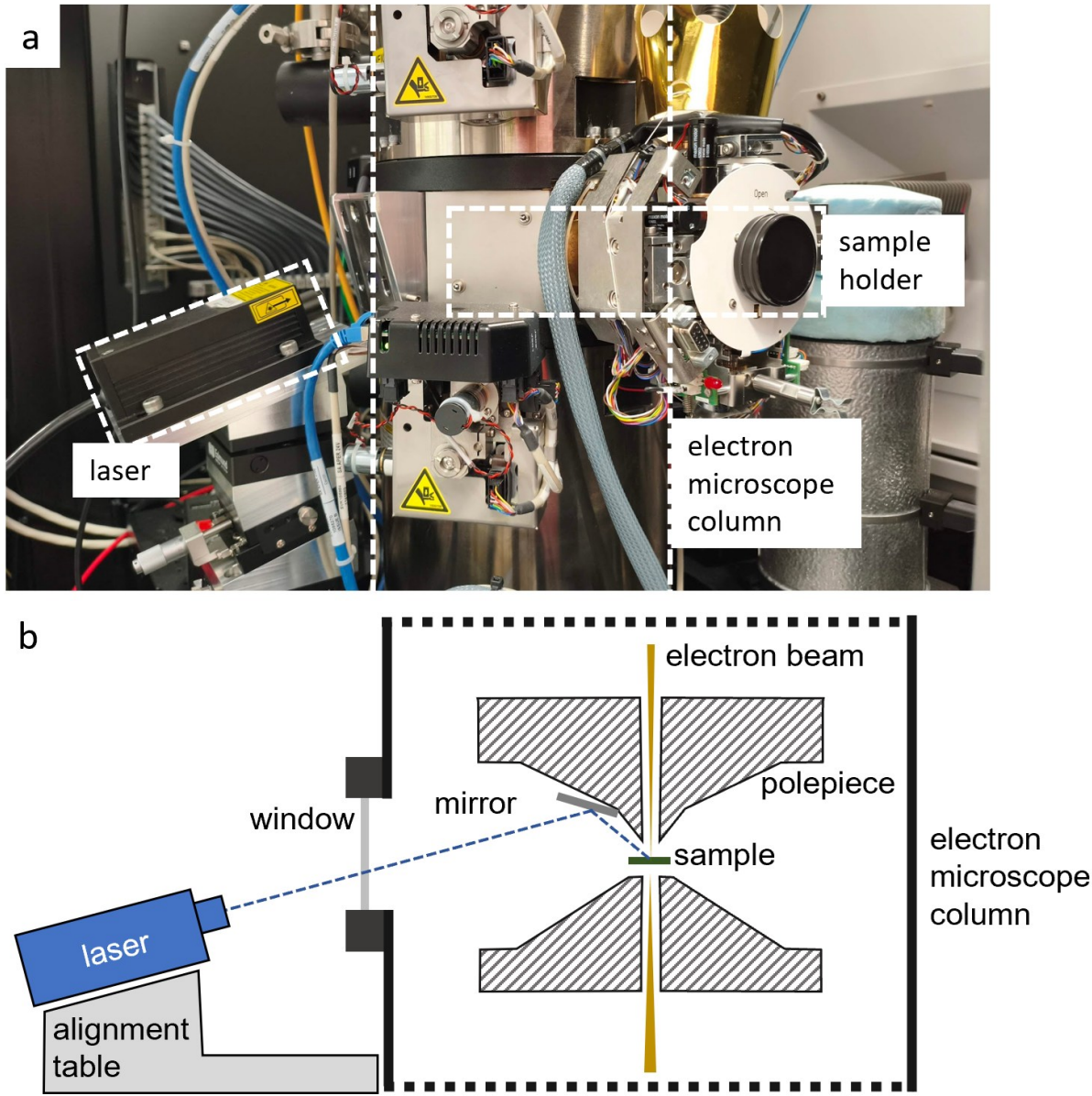

**Figure S6**. (a) Photograph of the laser illumination system in TEM. (b) Schematic cross section of the microscope column showing the illumination pathway.

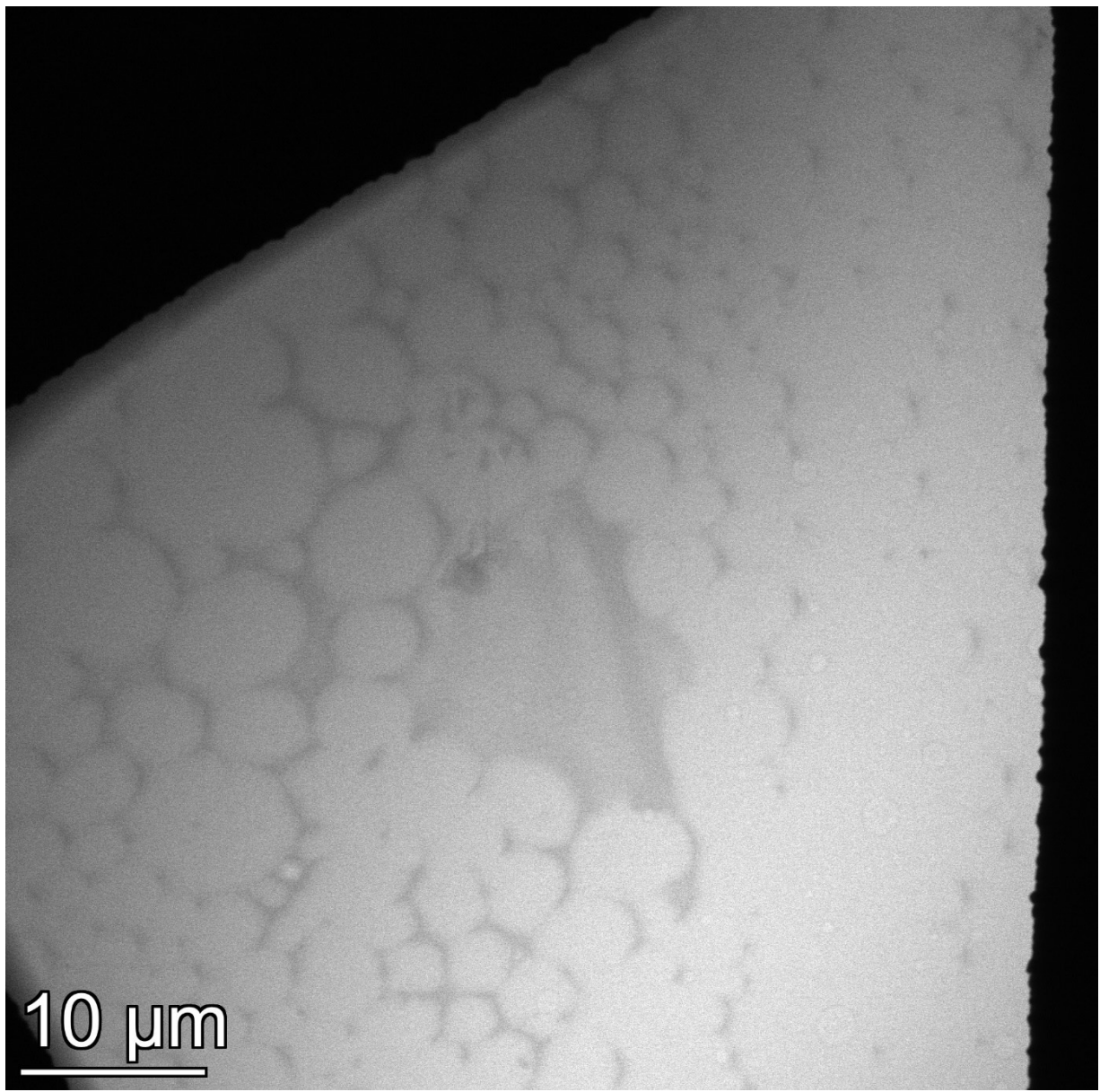

**Figure S7.** LP-STEM image of thick SmA islands formed upon chloroform evaporation during the sample preparation described in Methods section in the main article.